\documentclass{aa}  

\usepackage{graphicx}
\usepackage{txfonts}
\usepackage{mathrsfs}
\usepackage[colorlinks,allcolors=blue]{hyperref}
\usepackage{breakurl}
\usepackage{subfigure}

\def \src {\mbox{AX\,J0049.4$-$7323}}
\def \xmm {\emph{XMM-Newton}}
\def \asca {\emph{ASCA}}
\def \inte {\emph{INTEGRAL}}
\def \cha {\emph{Chandra}}
\def \rxte {\emph{RXTE}}
\def \sw  {\emph{Swift}}
\begin{document}

   \title{In-depth study of  long-term variability in the X-ray emission
          of the Be/X-ray binary system AX\,J0049.4$-$7323}
   \titlerunning{Long-term variability of AX\,J0049.4$-$7323}

   \author{L. Ducci
          \inst{1,2}
          \and
          P. Romano\inst{3}
          \and
          C. Malacaria\inst{1}
          \and
          L. Ji\inst{1}
          \and
          E. Bozzo\inst{2}
          \and
          A. Santangelo\inst{1}
          }

   \institute{Institut f\"ur Astronomie und Astrophysik, Kepler Center for Astro and Particle Physics, Eberhard Karls Universit\"at, 
              Sand 1, 72076 T\"ubingen, Germany\\
              \email{ducci@astro.uni-tuebingen.de}
              \and
              ISDC Data Center for Astrophysics, Universit\'e de Gen\`eve, 16 chemin d'\'Ecogia, 1290 Versoix, Switzerland
              \and
              INAF -- Osservatorio Astronomico di Brera, via Bianchi 46, 23807 Merate (LC), Italy
             }

   \date{Received ...; accepted ...}

\abstract
  {\src\ is a Be/X-ray binary in the Small Magellanic Cloud hosting a $\sim 750$\,s pulsar which has been 
    observed over the last $\sim$17 years by several X-ray telescopes. 
    Despite numerous observations, little is known about its X-ray behaviour.
    Therefore, we coherently analysed archival \sw, \cha, \xmm, \rxte, and \inte\ 
    data, and we compared them with already published \asca\ data, 
    to study its X-ray long-term spectral and flux variability.
    \src\  shows a high X-ray variability, spanning more than three orders of magnitudes,
    from $L\approx 1.6 \times 10^{37}$\,erg\,s$^{-1}$ (0.3$-$8\,keV, $d=62$\,kpc)
    down to $L \approx 8 \times 10^{33}$\,erg\,s$^{-1}$.
    \rxte, \cha, \sw, and \asca\ observed, in addition to the expected enhancement of X-ray
    luminosity at periastron, flux variations by a factor of $\sim 270$ with 
    peak luminosities of $\approx 2.1 \times 10^{36}$\,erg\,s$^{-1}$ far from periastron.
    These properties are difficult to reconcile with the typical long-term variability
    of Be/XRBs, traditionally interpreted in terms of type I and type II outbursts.
    The study of \src\ is complemented with a spectral analysis of \sw, \cha, and \xmm\ data which showed a softening trend 
    when the emission becomes fainter, and an analysis of optical/UV data collected by 
    the UVOT telescope on board \sw.
    In addition, we measured a secular spin-up rate of $\dot{P}=(-3.00\pm0.12)\times 10^{-3}$\,s\,day$^{-1}$,
    which suggests that the pulsar has not yet achieved its equilibrium period.
    Assuming spherical accretion, we
    estimated an upper limit for the magnetic field strength of the pulsar of $\approx 3\times 10^{12}$\,G.
  }


   \maketitle
%

\section{Introduction} 
\label{sect intro}

\src\ was discovered by \citet{Ueno00} 
in \asca\ observations of the Small Magellanic Cloud (SMC). 
It hosts a neutron star (NS) 
with a pulse period of $\approx 750$\,s \citep{Yokogawa00}
and a Be star \citep{Edge03} of spectral type O9.5-B0.5\,III-V \citep{McBride08}.
A spectroscopic analysis of the optical counterpart of \src\
showed a double-peaked H$\alpha$ emission profile, suggesting that
the circumstellar disc was observed  through the plane of rotation \citep{Edge03}.
\citet{Laycock05} and \citet{Cowley03} detected a periodicity of $\approx 395$\,d
in X-ray and optical bands interpreted as the orbital period of the NS.
\citet{Schmidtke13} combined MACHO and OGLE data to 
refine the optical period and ephemeris to $P_{\rm orb}=393.1 \pm 0.4$\,d
and $T_0=52979.5\pm 4$\,MJD.
\citet{Coe04} noticed that the two \rxte\ outbursts 
(exceeding $3\times 10^{37}$\,erg\,s$^{-1}$ in 3$-$10\,keV, assuming
a distance of $d=62$\,kpc)
used by \citet{Laycock05} to estimate the orbital period were
synchronized with the optical outbursts.
\citet{Galache08} reported further \rxte/PCA detections of 
\src\ in outburst, the brightest of which lasted three weeks.
\src\ was also detected by \inte\ during an outburst that reached
an X-ray luminosity $\gtrsim 10^{37}$\,erg\,s$^{-1}$ \citep{Coe10}.
\src\ was also detected three times by \asca. In all these cases,
it was observed far from the optical outbursts and it showed an
X-ray luminosity of $5\times 10^{35}$\,erg\,s$^{-1}$ \citep{Coe04}.
An analysis of archival \emph{ROSAT} and \emph{Einstein} data
showed \src\ at a relatively low luminosity level 
($\leq 5\times 10^{35}$\,erg\,s$^{-1}$) for over 20 years \citep{Yokogawa00}.
As noted by \citet{Coe04}, the three \asca\ detections at high X-ray luminosities
were difficult to explain in the context of the typical variability displayed 
by Be/X-ray binaries (Be/XRBs).
In general, Be/XRBs are transient X-ray sources 
with high eccentric orbits ($e \gtrsim 0.3$; \citealt{Reig11}).
The bright X-ray events characterizing the X-ray variability of Be/XRBs
are schematically divided into two groups, called type I and type II outbursts.
Type I outbursts are periodic (or quasi-periodic) events, occurring in correspondence with the
periastron. They last for a small fraction of the orbit
(0.2$-$0.3\,$P_{\rm orb}$) and their maximum luminosity is
typically of the order of $\lesssim10^{37}$\,erg\,s$^{-1}$.
They exceeds at least two orders of magnitude with respect to the quiescent
state, when the luminosity is $\lesssim 10^{35}$\,erg\,s$^{-1}$.
Type II outbursts are not periodic and can peak at any orbital phase.
They are brighter than type I outbursts and can reach X-ray luminosities of 
$\approx 10^{38}$\,erg\,s$^{-1}$.
They last for a large fraction of the orbit or for several orbital
periods \citep{Reig11}.
Tidal and resonant interactions of the NS with the circumstellar disc can lead
to the truncation of the latter at a radial distance which depends on the
viscosity of the disc \citep{Okazaki01, Negueruela01}.
In binary systems with low eccentricity ($e\lesssim 0.2$), the gap size
between the outer radius of the truncated disc and the orbit is so large
that no type I outbursts occur. These systems show only type II outbursts
and only occasionally type I outbursts, when the disc is strongly disturbed \citep{Reig11}.
In binary systems with high eccentricity ($e\gtrsim 0.6$) the disc truncation
is not efficient. At every periastron passage, the mass accretion rate is large
enough to cause regular type I outbursts \citep{Okazaki01}.
For systems with moderate eccentricity, the situation is more complicated.
The global properties of their X-ray variability depend more closely on
the orbital separation and viscosity properties of the circumstellar disc
\citep{Okazaki01}.
The \asca\ detections at luminosities of $5\times 10^{35}$\,erg\,s$^{-1}$ 
could not be classified as type I because they were
not observed at periastron and they were not bright enough for type II \citep{Coe04}.
The observations in hand of \citet{Coe04} were insufficient to draw further
conclusions about the anomalous X-ray behaviour shown by \src. 
Therefore, the aim of this work is to provide a detailed analysis of 
the largest possible set of X-ray data
to infer more information about the main properties of this system
through a global study of the X-ray long-term flux and spectral variability.
We coherently analysed \sw\ (XRT and UVOT), \cha, \xmm,
\rxte, and \inte\ data
(collected over the last $\sim$17 years and 
corresponding to a total exposure time of $\sim 5.86$\,Ms; Sect. \ref{sect. data analysis}).
The results are presented in Sect. \ref{sect. results} 
and discussed in Sect. \ref{sect. discussion}.

\section{Reduction and data analysis}
\label{sect. data analysis}

\subsection{Swift}

The SMC has been observed repeatedly with {\it Swift} \citep[][]{Gehrels2004}
since its launch; in particular, starting on  2016 June 8, the SMC has been observed as part of the
 {\it Swift} survey of the Small Magellanic Cloud \citep[][]{Kennea2016:atel9299_smcsurvey}. 
Therefore,  we collected all  observations in which the source was within the field of view of the
 narrow-field instruments, the X-ray Telescope \citep[XRT, ][]{Burrows2005:XRT}  
and the UV/Optical Telescope \citep[UVOT, ][]{Roming2005:UVOT}.
The observations are listed in Table \ref{xrtlog}. 

The XRT data were uniformly processed and analysed using the standard software 
({\sc FTOOLS}\footnote{\href{https://heasarc.gsfc.nasa.gov/ftools/ftools_menu.html}{https://heasarc.gsfc.nasa.gov/ftools/ftools\_menu.html.  } } v6.20), 
calibration (CALDB\footnote{\href{https://heasarc.gsfc.nasa.gov/docs/heasarc/caldb/caldb_intro.html}{https://heasarc.gsfc.nasa.gov/docs/heasarc/caldb/caldb\_intro.html.}}  20170501), 
and methods. 
We used the task {\sc xrtpipeline} (v0.13.3) to process and filter the XRT data, and extracted 
source events from a circular region with a radius of 10 pixels 
(1 pixel corresponds to $2.36$\arcsec). Background events were extracted  in most cases
from an annular region with an inner radius of 30  pixels and an external radius of 70 pixels 
centred  at the source position, the exception being when the source was close to the edge of the FOV; in this case  we used a circular region (with a radius of 50 to 70 pixels). 
The data are not affected by pile-up.
The XRT light curve was corrected for PSF losses and vignetting and background subtracted. 
For the spectral analysis,
we extracted events in the regions described above.
Then, we used xrtmkarf to generate ancillary response 
files that account for different extraction regions, vignetting, 
and PSF corrections. Spectra were extracted in each individual 
observation and  in several datasets combined. 
All were binned at 1 count per bin and fit adopting Cash \citep[][]{Cash79} statistics.

When no detection was achieved, the corresponding 3\,$\sigma$ upper limit 
on the X-ray count rate was estimated by using  the tasks {\it sosta} and {\it uplimit} within {\sc XIMAGE}  
(with the background calculated in the neighbourhood of the source position) 
and the Bayesian method for low-count experiments adapted from \citet[][]{KraftBurrowsNousek1991}.

UVOT observed the target simultaneously with XRT. 
As the observations were obtained from the {\it Swift} 
archive, there is no uniformity of filter usage in the UVOT data. 
The data analysis was performed using the {\it uvotimsum} and 
{\it uvotsource} tasks included in {\sc FTOOLS}. The {\it uvotsource} 
task calculates the magnitude of the source through aperture photometry within
a circular region centred on the best source position and applies the required corrections 
related to the specific detector characteristics. 
We adopted a circular region with a radius of 5\,\arcsec\ for the photometry of the 
different sources. The background was evaluated in all cases by using circular regions 
with a radius of 10\,\arcsec.  
For the magnitude uncertainties, we added in quadrature the systematic errors provided
by {\sc uvotsource}.

\begin{table*}  
 \begin{center}         
 \caption{Summary of the \sw/XRT observations. XRT was set in PC mode in all the observations listed below.
          Observations for which it was possible to extract meaningful X-ray spectra and fluxes are indicated with a star.
          Observations with the source within the UVOT field of view are indicated 
          with the name of the filter used in the observation.}         
 \label{xrtlog}         
 \resizebox{\columnwidth+\columnwidth}{!}{
\begin{tabular}{llllllllllll} 
 \hline 
 \hline 
 \noalign{\smallskip} 
 Sequence   & X-ray  &  UVOT    &Start time  (UT)  & End time   (UT) & Exposure  & Sequence   &  X-ray  &   UVOT    & Start time  (UT)  & End time   (UT) & Exposure \\ 
            &analysed& analysed &                  &                 &   (s)     &            & analysed&  analysed &                  &                 &      (s)   \\
  \noalign{\smallskip} 
 \hline 
 \noalign{\smallskip} 
00037787001& *       & M2 W1 W2 &       2008-08-18 02:52:37     &       2008-08-18 18:59:56        &       3079    & 00048739013&         &                &       2016-08-31 09:22:09        &       2016-08-31 09:23:09     &       43                  \\
00031428001& *       &          &       2009-06-21 01:24:19     &       2009-06-21 03:04:54        &       552      & 00048724014&         &               &       2016-09-21 03:17:39        &       2016-09-21 03:18:44     &       58                  \\
00090522001& *       & M2       &       2010-12-12 02:58:33     &       2010-12-12 12:30:58        &       9929     & 00048739014&         &               &       2016-09-21 03:39:49        &       2016-09-21 03:40:47     &       55                  \\
00090522002& *       &          &       2011-02-13 17:49:56     &       2011-02-14 19:44:56        &       6359     & 00048724015&         &               &       2016-09-28 09:11:41        &       2016-09-28 09:12:41     &       48                  \\
00090522003& *       &          &       2011-02-15 00:21:13     &       2011-02-15 21:38:58        &       1462     & 00048724016&         & W1            &       2016-10-05 11:25:51        &       2016-10-05 11:26:49     &       38                  \\
00090522004& *       &          &       2011-02-16 07:10:52     &       2011-02-16 18:16:56        &       2700     & 00048739016&         &               &       2016-10-05 11:47:55        &       2016-10-05 11:48:53     &       40                  \\
00090522005& *       &          &       2011-04-29 01:00:08     &       2011-04-30 23:46:58        &       9982     & 00048724017&         &               &       2016-10-12 08:02:59        &       2016-10-12 08:03:52     &       48                  \\
00040442001& *       &          &       2011-06-30 04:25:01     &       2011-06-30 07:52:58        &       2259     & 00048739017&         &               &       2016-10-12 10:48:53        &       2016-10-12 10:49:51     &       58                  \\
00040462001& *       &          &       2011-08-19 05:28:07     &       2011-08-19 15:11:56        &       562      & 00048724018&         &               &       2016-10-19 10:28:48        &       2016-10-19 10:29:51     &       25                  \\
00040439001& *       &          &       2011-08-20 02:19:53     &       2011-08-20 18:27:58        &       1908     & 00048739018&         &               &       2016-10-19 11:58:03        &       2016-10-19 11:59:03     &       55                  \\
00040440001& *       & M2 W1 W2 &       2011-08-20 13:33:54     &       2011-08-20 23:16:56        &       1637     & 00048724019&         &               &       2016-10-25 11:35:40        &       2016-10-25 11:36:40     &       53                  \\
00040440002& *       & M2 W1 W2 &       2011-08-21 21:40:40     &       2011-08-21 21:44:56        &       243      & 00048739019&         &               &       2016-10-25 21:07:21        &       2016-10-25 21:09:19     &       108                 \\
00032075002& *       &          &       2011-08-24 00:45:07     &       2011-08-24 23:34:56        &       6253     & 00048724020&         &               &       2016-10-26 03:24:19        &       2016-10-26 03:25:22     &       58                  \\
00032080001& *       & U        &       2011-08-24 17:02:34     &       2011-08-24 18:47:49        &       2941     & 00048739020&         &               &       2016-10-26 04:59:44        &       2016-10-26 05:00:44     &       25                  \\
00032075003& *       &          &       2011-08-25 06:58:37     &       2011-08-25 23:38:57        &       10531   & 00048724021&         &                &       2016-11-09 15:28:28        &       2016-11-09 15:29:21     &       38                  \\
00040462002& *       &          &       2011-08-26 18:43:23     &       2011-08-27 04:31:57        &       983     & 00048739021&         &                &       2016-11-09 15:50:24        &       2016-11-09 15:51:24     &       48                  \\
00040462003& *       &          &       2011-08-28 02:49:10     &       2011-08-29 22:16:54        &       1113    & 00048724022&         &                &       2016-11-16 07:00:10        &       2016-11-16 07:01:10     &       53                  \\
00040439002& *       &          &       2011-08-29 01:10:58     &       2011-08-29 01:21:57        &       639     & 00048739022&         &                &       2016-11-16 08:36:18        &       2016-11-16 08:37:18     &       43                  \\
00040440003& *       & M2 W1 W2 &       2011-08-31 09:11:41     &       2011-08-31 09:25:56        &       835     & 00048724023&         &                &       2016-11-23 15:57:43        &       2016-11-23 15:58:26     &       20                  \\
00040439003& *       &          &       2011-08-31 22:02:14     &       2011-08-31 22:10:58        &       501     & 00048739023&         &                &       2016-11-23 17:27:14        &       2016-11-23 17:28:16     &       58                  \\
00032194001& *       &          &       2011-11-29 09:59:47     &       2011-11-29 10:15:57        &       948     & 00048724024&         &                &       2016-11-30 07:31:42        &       2016-11-30 07:32:35     &       38                  \\
00034071001& *       & U        &       2015-09-24 02:21:22     &       2015-09-24 05:27:20        &       444     & 00048725024&         &                &       2016-11-30 08:47:34        &       2016-11-30 08:48:29     &       45                  \\
00088083001& *       &          &       2017-03-12 01:58:46     &       2017-03-13 19:33:52        &       7226     & 00048724025&         &               &       2016-12-07 07:54:38        &       2016-12-07 07:55:41     &       45                  \\
00088083002& *       & M2       &       2017-03-14 00:03:15     &       2017-03-14 06:35:53        &       7311     & 00048739025&         &               &       2016-12-07 08:16:44        &       2016-12-07 08:17:44     &       45                  \\
00088083003& *       & M2       &       2017-03-15 01:30:52     &       2017-03-15 11:40:52        &       7304     & 00048724026&         &               &       2016-12-14 05:46:44        &       2016-12-14 05:47:44     &       40                  \\
00048724001&         & W1       &       2016-06-08 03:50:57     &       2016-06-08 03:51:57        &       38       & 00048739026&         &               &       2016-12-14 07:18:47        &       2016-12-14 07:19:49     &       50                  \\
00048739001&         & W1       &       2016-06-08 06:51:07     &       2016-06-08 06:52:08        &       50       & 00048724027&         &               &       2016-12-29 08:21:11        &       2016-12-29 08:22:12     &       58                  \\
00048724002&         &          &       2016-06-16 02:58:52     &       2016-06-16 02:59:50        &       45      & 00048739027&         &                &       2016-12-29 09:57:26        &       2016-12-29 09:58:24     &       38                  \\
00048739002&         &          &       2016-06-16 04:29:12     &       2016-06-16 04:30:12        &       50      & 00048724028&         &                &       2017-01-05 14:04:49        &       2017-01-05 14:05:52     &       50                 \\
00048724003&         &          &       2016-06-24 03:38:32     &       2016-06-24 03:39:32        &       48      & 00048725028&         &                &       2017-01-05 14:06:15        &       2017-01-05 14:07:20     &       15                 \\
00048724004&         &          &       2016-06-28 12:45:00     &       2016-06-28 12:46:03        &       53      & 00048739028&         &                &       2017-01-05 15:31:55        &       2017-01-05 15:32:58     &       45                 \\
00048739004&         &          &       2016-06-28 14:21:43     &       2016-06-28 14:22:43        &       53      & 00048724029&         &                &       2017-01-11 08:32:58        &       2017-01-11 08:34:00     &       63                 \\
00048724005&         &          &       2016-07-06 12:21:15     &       2016-07-06 12:22:10        &       53      & 00048739029&         &                &       2017-01-11 08:55:05        &       2017-01-11 08:56:03     &       55                 \\
00048739005&         &          &       2016-07-06 13:58:44     &       2016-07-06 13:59:39        &       55      & 00048724030&         &                &       2017-01-18 08:09:37        &       2017-01-18 08:10:40     &       55                 \\
00048724006&         &          &       2016-07-10 11:32:55     &       2016-07-10 11:33:53        &       50      & 00048739030&         &                &       2017-01-18 14:43:56        &       2017-01-18 14:44:59     &       63                 \\
00048739006&         &          &       2016-07-10 11:54:55     &       2016-07-10 11:56:00        &       60      & 00048724031&         &                &       2017-01-25 05:39:04        &       2017-01-25 05:40:04     &       45                 \\
00048739007&         &          &       2016-07-15 17:43:27     &       2016-07-15 17:44:25        &       45      & 00048739031&         &                &       2017-01-25 10:28:48        &       2017-01-25 10:29:53     &       45                 \\
00048724007&         &          &       2016-07-15 16:01:34     &       2016-07-15 16:02:27        &       43      & 00048724032&         &                &       2017-02-01 02:03:17        &       2017-02-01 02:04:20     &       50                 \\
00048724008&         & M2       &       2016-07-29 17:06:29     &       2016-07-29 17:07:34        &       55      & 00048739032&         & W1             &       2017-02-01 03:35:31        &       2017-02-01 03:36:34     &       43                 \\
00048739008&         &          &       2016-07-29 18:43:22     &       2016-07-29 18:44:22        &       50      & 00048724033&         & W1             &       2017-02-22 05:16:52        &       2017-02-22 05:17:50     &       40                 \\
00048724009&         &          &       2016-08-03 03:55:41     &       2016-08-03 03:56:46        &       63      & 00048724034&         & W1             &       2017-03-08 02:41:19        &       2017-03-08 02:42:19     &       53                 \\
00048739009&         &          &       2016-08-03 05:34:08     &       2016-08-03 05:35:13        &       45      & 00048724035&         &                &       2017-03-14 06:59:50        &       2017-03-14 07:00:53     &       60                 \\
00048724010&         &          &       2016-08-10 01:57:56     &       2016-08-10 01:58:56        &       48      & 00048739035&         &                &       2017-03-14 08:25:03        &       2017-03-14 08:25:58     &       43                 \\
00048739010&         &          &       2016-08-10 05:13:19     &       2016-08-10 05:14:21        &       53      & 00048724036&         &                &       2017-03-22 04:23:21        &       2017-03-22 04:24:14     &       35                 \\
00048724011&         & W1       &       2016-08-17 08:55:58     &       2016-08-17 08:56:53        &       43      & 00048739036&         &                &       2017-03-22 07:29:17        &       2017-03-22 07:30:20     &       63                 \\
00048739011&         &          &       2016-08-17 10:34:53     &       2016-08-17 10:35:48        &       40      & 00048739033&         &                &       2017-02-22 05:38:49        &       2017-02-22 05:39:54     &       45                 \\
00048724012&         & W1       &       2016-08-24 10:03:49     &       2016-08-24 10:04:47        &       45      & 00048724037&         &                &       2017-03-29 17:57:20        &       2017-03-29 17:58:25     &       5                  \\
00048739012&         & W1       &       2016-08-24 11:33:56     &       2016-08-24 11:34:57        &       48      & 00048739037&         &                &       2017-03-30 00:06:23        &       2017-03-30 00:07:19     &       35                 \\
00048724013&         & W1       &       2016-08-31 07:55:06     &       2016-08-31 07:55:56        &       30      &            &         &                &                               &                               &                          \\    
  \noalign{\smallskip}
  \hline
  \end{tabular}
}
  \end{center}
  \end{table*}

\subsection{Chandra}
\label{sec:chandra}

\cha\ data have been explored to analyse the long-term behaviour of \src.
We extracted images for each of the \cha\ observations in which the source 
was detected within the ACIS field of view (see Table \ref{Table spectra}).
We applied the latest \cha\ calibration files to each image and we reduced the data 
with the \cha\ Interactive Analysis of Observations software package (CIAO, version $4.9$).
Source fluxes were extracted with the CIAO tool \textit{srcflux} \citep{Glotfelty+14}. 
We extracted source events from circular regions with radius of about $3''-5''$, 
while the background was extracted from an annular region 
around the source, with internal radius as large as twice that of the 
source region and outer radius equal to about $5$ times the source radius.
The extraction energy band was constrained to $0.3-8\,$keV.
Due to the low count rates detected from this source 
and its peripheral position on the ACIS detectors (which smears the detection region), 
pile-up effects are negligible (pile-up fraction generally $<1\%$).
\cha\ spectra have been extracted using the CIAO \textit{specextract} tool
and analysed using XSPEC (ver. 12.9.1, \citealt{Arnaud96}).
Spectral channels were rebinned to contain at least 25 photons per energy bin.

\subsection{XMM-Newton}

\xmm\ observed three times the field around \src:
2000-10-15 15:09:54 (UTC), 2007-04-11 19:37:46,
and 2009-10-03 05:08:04 (see Table \ref{Table spectra}).
The data analysis of each observation was performed
through the \xmm\ Science Analysis System (SAS) sofware
(version 15.0.0). Time intervals affected by high background
were excluded, and calibrated event lists for pn, MOS1, and MOS2
were produced using the {\it epproc} and {\it emproc} tasks.
For the pn data we used single- and double-pixel events (PATTERN$\leq$4),
while for the MOS data, single-pixel to quadruple-pixel events
(PATTERN$\leq$12) were used.
For each observation we extracted in each detector the source events
in circular regions centred on the target, using appropriate radii,
according to the brightness of the source and its location in the detector.
Background events were accumulated from source-free circular regions 
far from the point spread function of the source.
We extracted the spectra of the first two observations.
Spectral channels  were rebinned to contain at least 25 photons per energy bin.
For each observation, we fitted all the available EPIC spectra simultaneously
using XSPEC.

\subsection{RXTE}
\label{subs rxte}

We analysed \rxte/PCA corresponding to the detections reported
in \citet{Yang17} and \citet{Klus14}.
We used Standard\,2 mode PCA data and, for each observation, we extracted
light curves  in the 3$-$10\,keV energy range, with time bins of 16\,s.
Data were screened with standard criteria. Corrections for deadtime were applied.
We selected events in all the active PCUs and layers.
The barycentric correction was applied using {\it faxbary}.

\subsection{INTEGRAL}
\label{subs integral}

We used data collected by the detector ISGRI
of the coded-mask telescope IBIS 
(\citealt{Lebrun03}; \citealt{Ubertini03}).
ISGRI operates in the energy band $\sim 15-400$~keV.
\inte\ observations consist of pointings called science windows (ScWs).
We performed the data reduction using the Off-line Science Analysis (OSA) 10.2
software. We analysed the public data from 2003 July 23
to 2015 December 7 (52843.96$-$57363.26\,MJD) in which \src\ was within 12$^\circ$
of the centre of the IBIS/ISGRI field of view\footnote{At larger off-axis angle the response is not well known. 
See the \inte\ data analysis documentation:
\url{http://www.isdc.unige.ch/integral/analysis\#Documentation}}.
The total exposure time is $\approx 5.86$~Ms.
\emph{INTEGRAL} observations from 2008 November 11 to 2009 June 25
were already analysed and the results presented in \citet{Coe10}.
These data are re-analysed here to ensure uniformity.

\section{Results}
\label{sect. results}

\subsection{Spectral analysis} 
\label{sect spectra}

For \cha, \xmm, and \sw\ observations with sufficiently high statistics, we extracted X-ray 
spectra. 
We found that a simple absorbed power law model instead of a more complicated model, such as an absorbed cutoff power law,
can describe the spectra adequately well. This is due to the relatively poor statistics
and the high value of the e-folding energy of this source which,
according to \citet{Laycock05}, is $E_{\rm f}=14$\,keV, hence
outside of the energy range covered by our data.
For the absorption component, we adopted {\tt tbabs} in XSPEC.
The values of the spectral parameters are listed in Table \ref{Table spectra}.
The Galactic absorption in the direction of \src\ ranges from 
$N_{\rm H} \approx 4\times 10^{20}$\,cm$^{-2}$ to $N_{\rm H} \approx 4\times 10^{21}$\,cm$^{-2}$
\citep{Kalberla05,Dickey90}.
Due to this large uncertainty, we decided not to include a fixed foreground absorption
component in the spectral model.
The column densities resulting from the spectral fit 
(Table \ref{Table spectra}) are always slightly higher than   
the lower value of the foreground absorption reported above,
indicating a non-negligible intrinsic absorption around the source.
For observations with insufficient statistics to perform a spectral analysis,
we reported in Table \ref{Table spectra} the flux in the range 0.3$-$8\,keV obtained from the conversion
of the count rate, assuming an absorbed power law spectrum with 
$N_{\rm H}=4\times 10^{21}$\,cm$^{-2}$ and photon index of 0.8.
The results of the spectral analysis of the two \xmm\ observations 0110000101 and
0404680301 and the \cha\ observations 8479, 7156, and  8481 
are consistent, within the 90\% confidence level (c.l.) with those 
reported in \citet{Haberl04, Haberl08}, and \citet{Hong17}.
The observed X-ray fluxes are also consistent with those reported by \citet{Yang17}.
The results from the other observations are published here for the first time.
The slopes of the power law given in Table \ref{Table spectra} 
are typical of accreting pulsars.

We grouped the \cha\ observations of the source detected at the lowest
luminosity state (when $F_{\rm x} < 10^{-12}$\,erg\,cm$^{-2}$\,s$^{-1}$)
to perform a meaningful spectral analysis.
The resulting average spectrum can be fitted with 
an absorbed power law model with $N_{\rm H}=0.73{+0.26 \atop -0.24}10^{22}$\,cm$^{-2}$,
$\Gamma = 1.63{+0.28 \atop -0.26}$, $\chi^2_\nu=1.23$ (11 d.o.f.).
The power law slope is higher than in the observations 
where the source is detected at higher fluxes (Table \ref{Table spectra}).

   \begin{figure}
   \centering
   \includegraphics[bb=89 369 554 698, width=9cm]{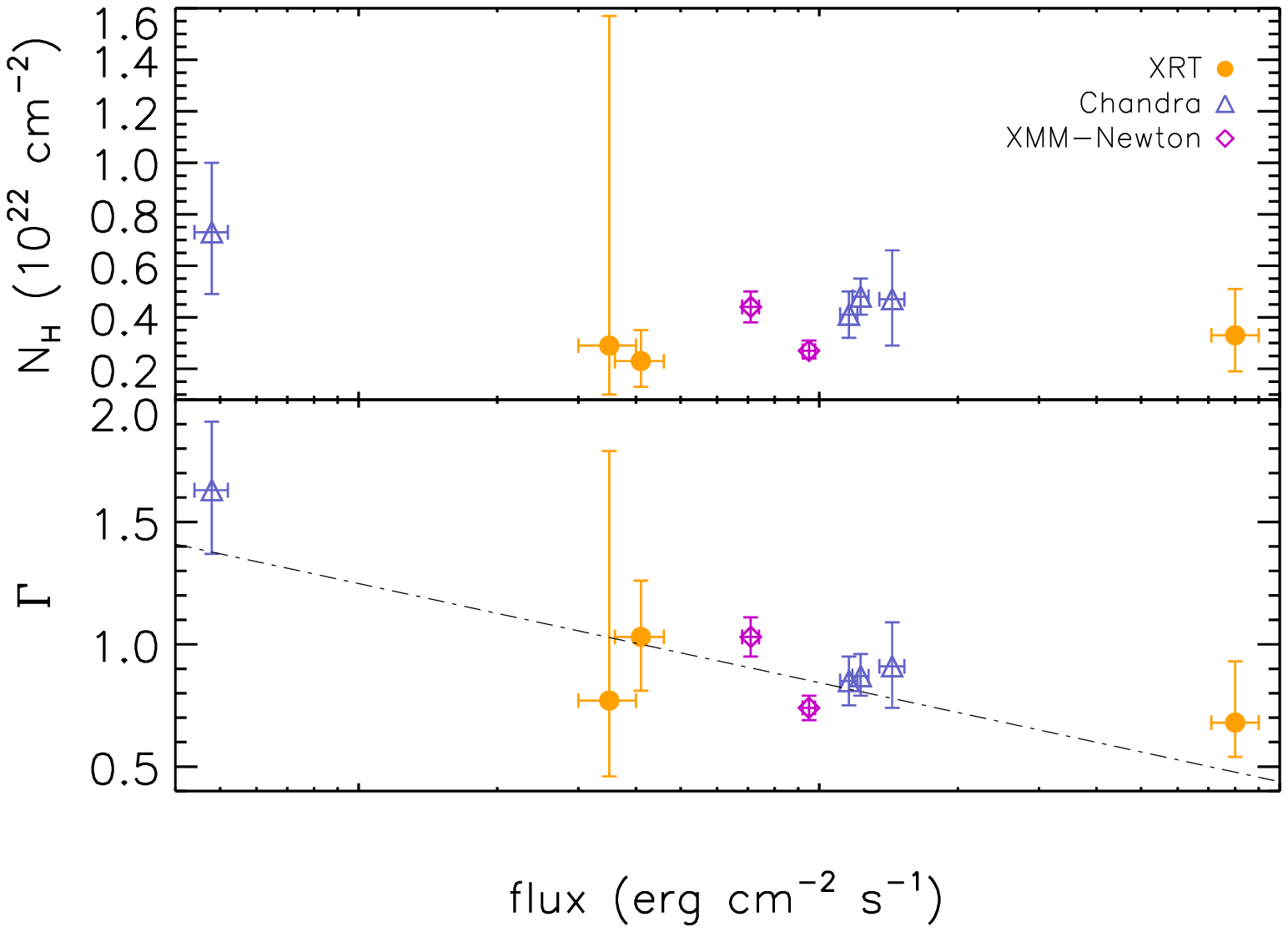}\\
   \includegraphics[bb=70 22 583 720, angle=-90, width=9cm]{cha_2xmm.ps}
      \caption{\emph{Top panel:} Column density and photon index as a function of the absorbed flux
                                 (errors quoted at 1$\sigma$ c.l.). The plotted values
               correspond to those listed in Table \ref{Table spectra}. The anti-correlation $\Gamma$ vs flux
               is emphasized by the dot-dashed line, which shows the best linear fit ($\Gamma$ vs $\log_{10} F_{\rm x}$).
               \emph{Bottom panel:} Confidence contours at 68\%, 95\%, and 99\%  in the $N_{\rm H}-\Gamma$
               plane of the \cha\ spectrum at the lowest luminosity level
               and the \xmm\ spectra.}
         \label{correlazioni}
   \end{figure}

The upper panel of Fig. \ref{correlazioni} shows the values of $N_{\rm H}$ and $\Gamma$
in Table \ref{Table spectra} as a function of the absorbed flux.
The plot shows that the spectral slope softens at lower fluxes
($\Gamma = (-4.0 \pm 1.7)\times\log_{10} F_{\rm x} -(0.4 \pm 0.14)$, errors quoted at 1$\sigma$ c.l.).
To quantify the significance of the anticorrelation $\Gamma$ versus flux,
we calculated the Pearson's linear coefficient $r$ and the 
null hypothesis probability $p$ in the $\log_{10} x - y$ space.
We found $r=-0.83$ and $p=0.5$\%. The parameter 
$N_{\rm H}$ does not show a significant correlation with  flux;
in fact, a linear fit of $N_{\rm H}$ versus $\log_{10} F_{\rm x}$ results in 
$N_{\rm H} \propto (0.17 \pm 0.22)\times\log_{10} F_{\rm x}$, where the error is quoted at $2\sigma$ c.l.
The slope is thus consistent with zero at $2\sigma$ c.l.
To highlight the significant spectral variability of \src, 
we show in Fig. \ref{correlazioni}, lower panel, the
68\%, 95\%, and 99\% confidence contours in the $N_{\rm H}-\Gamma$
plane of the average \cha\ spectrum at the lowest luminosity level
(the   `combined' spectrum in Table \ref{Table spectra})
and the \xmm\ spectra where the flux was up to $\sim$20 times larger.

\subsection{Timing analysis}
\label{subs timing}

\begin{figure}
   \centering
   \includegraphics[width=\columnwidth]{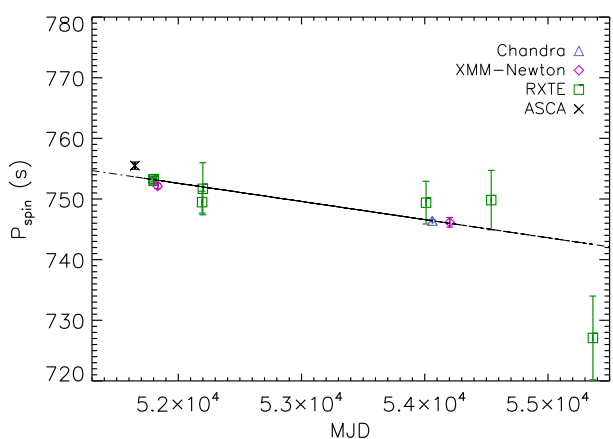}
      \caption{Spin period evolution of \src. The black line shows the best-fit long-term spin-up.
               Error bars indicate 1$\sigma$ uncertainties.}
         \label{spin evol}
   \end{figure}

We used the Lomb--Scargle periodogram technique \citep{Press89}
to search for periodicities in \xmm, \cha, and \rxte\ data.
\sw\ observations were too short compared to the spin period of \src\
to search for periodicities.
For \rxte\ data, we performed timing analysis in each observation and
in groups of observations within one day.
We searched for periodicities within a small window
ranging from 720\,s to 780\,s. We set the limits of this window on the basis of the spin period history
of the source reported by \citet{Yang17}.
We set a false alarm probability of detection of $99.9$\%.
The number of independent trial frequencies was calculated according to equation 13 of \citet{Horne86}.
We detected pulsations in the \xmm\ observations 0110000101 and 0404680301,
and in the \cha\ observations 8479, 7156, and 8481.
For these observations, our measurements of the spin period are in agreement with those reported
in  \citet{Haberl08} and \citet{Hong17}, hence hereafter we use the \cha\ and \xmm\ spin periods
published in those papers.
In the other \cha\ and \xmm\ observations we did not detect any pulsating signal
probably because of the insufficient statistics (see Sect. \ref{sect. gating mechanism}).
We detected \src\ in the \rxte\ observations reported in Table \ref{Table timing}.
Figure \ref{spin evol} shows the long-term spin period evolution of \src.
In addition to the spin period measurements obtained in this work, 
we included the spin period measured by \citet{Yokogawa00} using \asca\ ($P_\asca=755.5\pm 0.6$\,s).
Figure \ref{spin evol} shows a significant long-term spin-up of $\dot{P}=(-3.00\pm0.12)\times 10^{-3}$\,s\,day$^{-1}$.

The source does not show significant variability such as flares within each observation,
in agreement with previous findings obtained by \citet{Hong17}, \citet{Haberl08}, and \citet{Haberl04}, based on
\cha\ and \xmm\ data.

\begin{table}
\begin{center}
\caption{Measurements of the spin period and 0.3$-$8\,keV flux 
of \src\ in \rxte\ observations.}
\label{Table timing}
\resizebox{\columnwidth}{!}{
\begin{tabular}{lccc}
\hline
\hline
\noalign{\smallskip}
T$_{\rm start}$ & T$_{\rm stop}$ & $P_{\rm spin}$ &          $F_{\rm x}$               \\
\multicolumn{2}{c}{MJD}     &     (s)      & ($10^{-11}$\,erg\,$cm^{-2}$\,s$^{-1}$) \\  
\noalign{\smallskip}
\hline
\noalign{\smallskip}
51800.00   &  51800.70  &  $753.24\pm 0.42$  & $1.75 \pm 0.16$ \\
\noalign{\smallskip}
51800.86   &   51801.30 &   $753.00\pm 0.41$ & $1.62 \pm 0.16$   \\
\noalign{\smallskip}
51801.44   &   51802.28 &  $753.28\pm 0.31$  & $1.48 \pm 0.16$ \\
\noalign{\smallskip}
52193.617  &   52193.76 &   $749.48\pm 1.81$ & $3.50 \pm 0.16$  \\
\noalign{\smallskip}
52199.47   &   52199.57 &   $751.686\pm 4.3$ & $1.25 \pm 0.16$  \\
\noalign{\smallskip}
54010.46   &   54010.73 &   $749.398\pm 3.52$ & $0.46 \pm 0.16$  \\
\noalign{\smallskip}
54539.03   &   54539.14 &   $749.82\pm 4.9$  & $1.21 \pm 0.16$  \\
\noalign{\smallskip}
55363.906  &  55364.087 &   $727.09\pm 6.9$  & $0.43 \pm 0.16$  \\
\noalign{\smallskip}
\hline
\end{tabular}
}
\end{center}
\end{table}

   \begin{figure*}
   \centering
   \includegraphics[width=\columnwidth+\columnwidth]{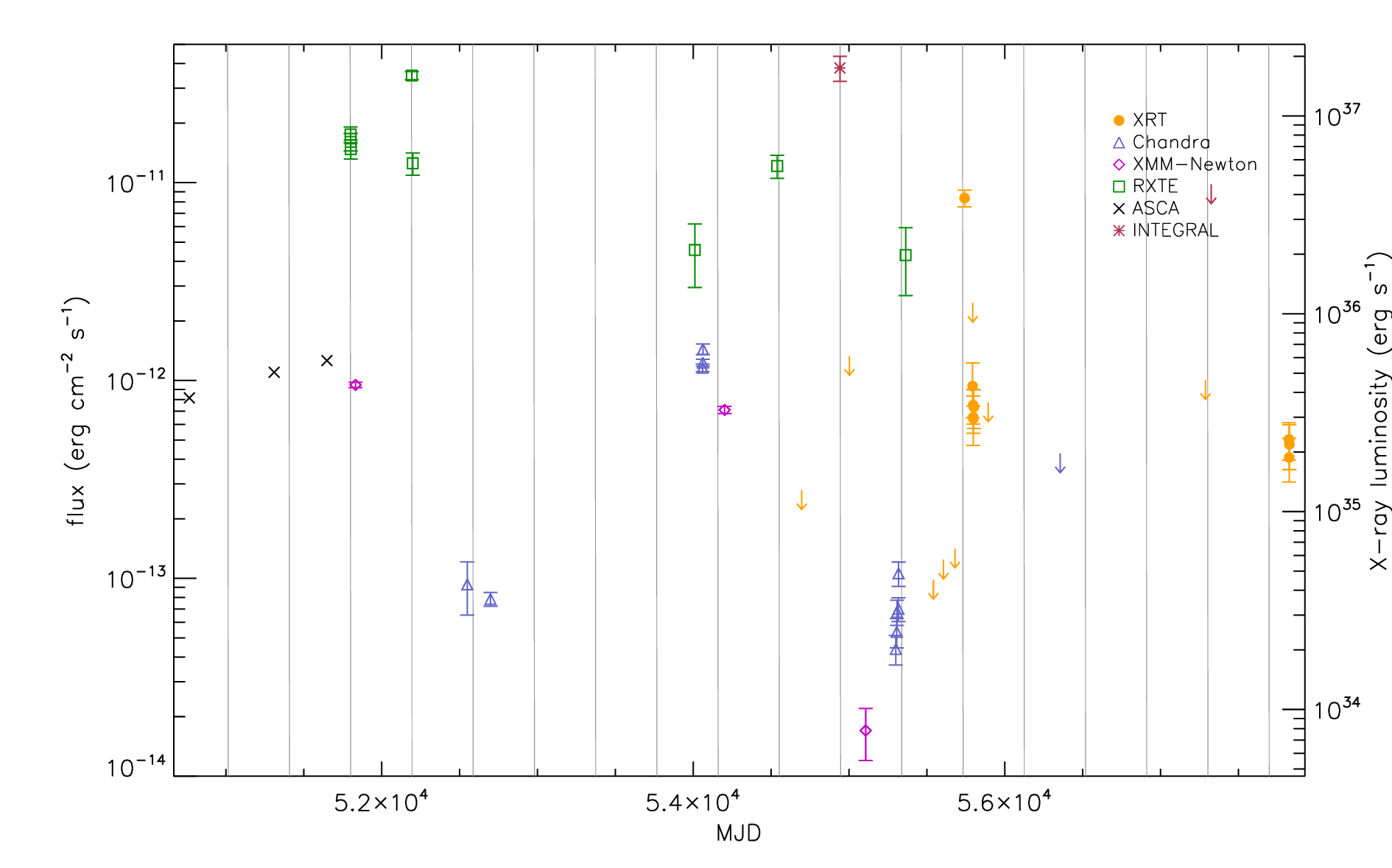}
      \caption{X-ray light curve (0.3$-$8\,keV) of \src\ obtained from \sw/XRT, \cha, \xmm, and \rxte\ data analysed in this work,
        and \asca\ data from \citet{Yokogawa00}. \inte\ points are in the energy range 20$-$40\,keV. 
        Downward arrows are used to indicate 3$\sigma$ upper limits.
               Grey vertical lines show the times of periastron passages
               according to the ephemeris calculated by \citet{Schmidtke13}.
               The right y-axis shows the absorbed $0.3-8$\,keV X-ray luminosity, assuming $d=62$\,kpc.
               Error bars indicate the 1$\sigma$ uncertainties.}
         \label{xray ogle}
   \end{figure*}

\subsection{Long-term flux variability}
\label{sect longterm}

Figure \ref{xray ogle} shows the long-term flux history of \src, from
2000 April 11 to 2017 March 15, based on \sw/XRT, \cha, \xmm, \emph{RXTE}/PCA,
\inte, and \asca\ data.
Each point, except for \inte, represents the average 0.3$-$8\,keV 
flux measured during the different observations.
For \inte, sky images of each pointing were generated in the energy band $20-40$~keV.
\src\ was never detected, being below the $5\sigma$ threshold of detection,
in the individual sky images of each pointing.
Then, for each \emph{INTEGRAL} revolution, we combined the individual images
to produce mosaic images. As already shown in \citet{Coe10},
\src\ is detected with a significance of 6.9$\sigma$ during the \emph{INTEGRAL} revolution
796, corresponding to the time interval 54941.368$-$54942.975\,MJD and orbital phase $\phi=0.992$.
The 20$-$40~keV is $F_{\rm x}=(3.8 \pm 0.6 )\times 10^{-11}$\,erg\,cm$^{-2}$\,s$^{-1}$,
corresponding to a luminosity of $(1.7 \pm 0.3) \times 10^{37}$\,erg\,s$^{-1}$.
\src\ was never detected in the other revolution-based mosaic images.
It has been observed with a long exposure (429\,ks)
during the periastron passage $\sim$57316$-$57335 MJD.
We measured a $20-40$~keV $3\sigma$ flux upper limit of $9.8\times 10^{-12}$\,erg\,cm$^{-2}$\,s$^{-1}$.

\begin{figure}
   \centering
    \includegraphics[bb=76 373 662 763,clip,width=\columnwidth]{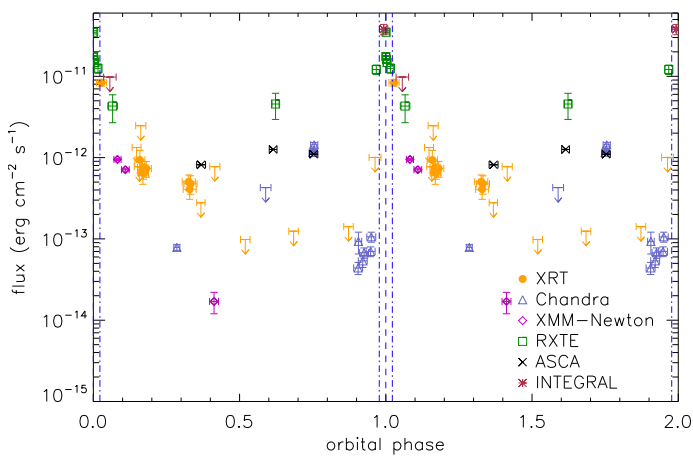}
    \caption{\sw, \cha, \xmm, and \rxte\ 0.3$-$8\,keV fluxes folded at the orbital period of \src.
      \inte\ points are in the energy range 20$-$40\,keV. 
      Downward arrows are used to indicate 3$\sigma$ upper limits.
               Vertical dashed lines and dot-dashed lines show the position of the periastron passages
               and their uncertainties, respectively.}
         \label{xray folded}
   \end{figure}

For \rxte\ data, we used equation 5 of \citet{Yang17} to calculate,
from each pulsation amplitude and corresponding uncertainty,
the $0.3-8$\,keV flux of the source.
We converted the flux from units counts/PCU/s to erg\,cm$^{-2}$\,s$^{-1}$
using PIMMS and assuming an absorbed power law model with $N_{\rm H}=4\times 10^{21}$\,cm$^{-2}$
and a photon index of $0.8$. For the pulsed fraction, we assumed, in accordance with \citet{Yang17}
and the typical values of \src\ \citep{Hong17}, $p_{\rm f}=0.4$.
The \rxte\ fluxes depend on the spectral and pulsed fraction parameters reported above,
  which are variable.
  Therefore, \rxte\ fluxes could be subject to large uncertainties.
  By varying the spectral parameters and $p_{\rm f}$  according to Table \ref{Table spectra}
  and  \citet{Hong17} and \citet{Haberl08},
  we find that the \rxte\ fluxes are known with a relative uncertainty of $\sim 50$\%.
Figure \ref{xray ogle} shows that \rxte\ data have a systematically  higher flux
than the observations from other instruments, also when the
  relative uncertainties of $\sim 50$\% of \rxte\ fluxes are taken into account.
This is a selection bias caused by the larger
number of \rxte/PCA observations with \src\ in the field of view (391, \citealt{Yang17})
compared to the low number of observations from other telescopes, together with the relatively low
sensitivity of \rxte/PCA (flux threshold: $\approx 4\times 10^{-12}$\,erg\,cm$^{-2}$\,s$^{-1}$ in 
the 2$-$10\,keV energy band; \citealt{Jahoda96}).
In addition to \sw, \xmm, \cha, and \rxte, we plotted the \asca\ fluxes reported in \citet{Yokogawa00}
and converted to the 0.3$-$8\,keV energy range  assuming an absorbed power law spectrum
with the parameters found by \citet{Yokogawa00} during the \asca\ observations.
\inte\ data of Fig. \ref{xray folded} are instead in the 20$-$40\,keV energy range.

\src\ shows a high X-ray variability, spanning more than three orders of magnitude,
from $\sim 1.7 \times 10^{-14}$\,erg\,cm$^{-2}$\,s$^{-1}$ 
(corresponding to a luminosity of $L_{\rm x}\sim8\times 10^{33}$\,erg\,s$^{-1}$ in the energy range 0.3$-$8\,keV)
to $\sim 3.5 \times 10^{-11}$\,erg\,cm$^{-2}$\,s$^{-1}$ ($L_{\rm x}\sim1.6\times 10^{37}$\,erg\,s$^{-1}$).
We found that, in addition to \asca\ \citep{Coe04},
 \sw, \rxte, and \cha\ also observed \src\ at orbital phases
not coinciding with the periastron during relatively 
high luminosity states ($L_{\rm x} > 2 \times 10^{35}$\,erg\,s$^{-1}$).

Figure \ref{xray folded} shows the X-ray light curve  folded at the orbital period
of \src\ (we adopted the ephemeris calculated by \citealt{Schmidtke13}).
At periastron ($0.02\lesssim \phi\lesssim 0.98$) the flux is on average higher than elsewhere  
and varies by a factor of $\sim$4,
from $F_{\rm x}\approx 8\times 10^{-12}$\,erg\,cm$^{-2}$\,s$^{-1}$ ($L_{\rm x}\approx 3.7\times 10^{36}$\,erg\,s$^{-1}$)
to $F_{\rm x}\approx 3.5\times 10^{-11}$\,erg\,cm$^{-2}$\,s$^{-1}$ ($L_{\rm x}\approx1.6\times 10^{37}$\,erg\,s$^{-1}$).
Outside the periastron passage, 
the maximum variability factor is $\approx$270,
and the flux (luminosity) ranges from
$F_{\rm x}\approx 1.7 \times 10^{-14}$\,erg\,cm$^{-2}$\,s$^{-1}$ ($L_{\rm x}\approx 8 \times 10^{33}$\,erg\,s$^{-1}$) 
to  $F_{\rm x}\approx 4.6 \times 10^{-12}$\,erg\,cm$^{-2}$\,s$^{-1}$ ($L_{\rm x}\approx 2.1 \times 10^{36}$\,erg\,s$^{-1}$).
The light curve does not show a sinusoidal modulation. On the contrary, it is characterized (see also Fig. \ref{xray ogle}) by jumps in luminosity apparently uniformly
distributed across the orbit rather than being clustered at some particular orbital phase.

\subsection{UVOT}

\begin{figure}
   \centering
   \includegraphics[width=\columnwidth]{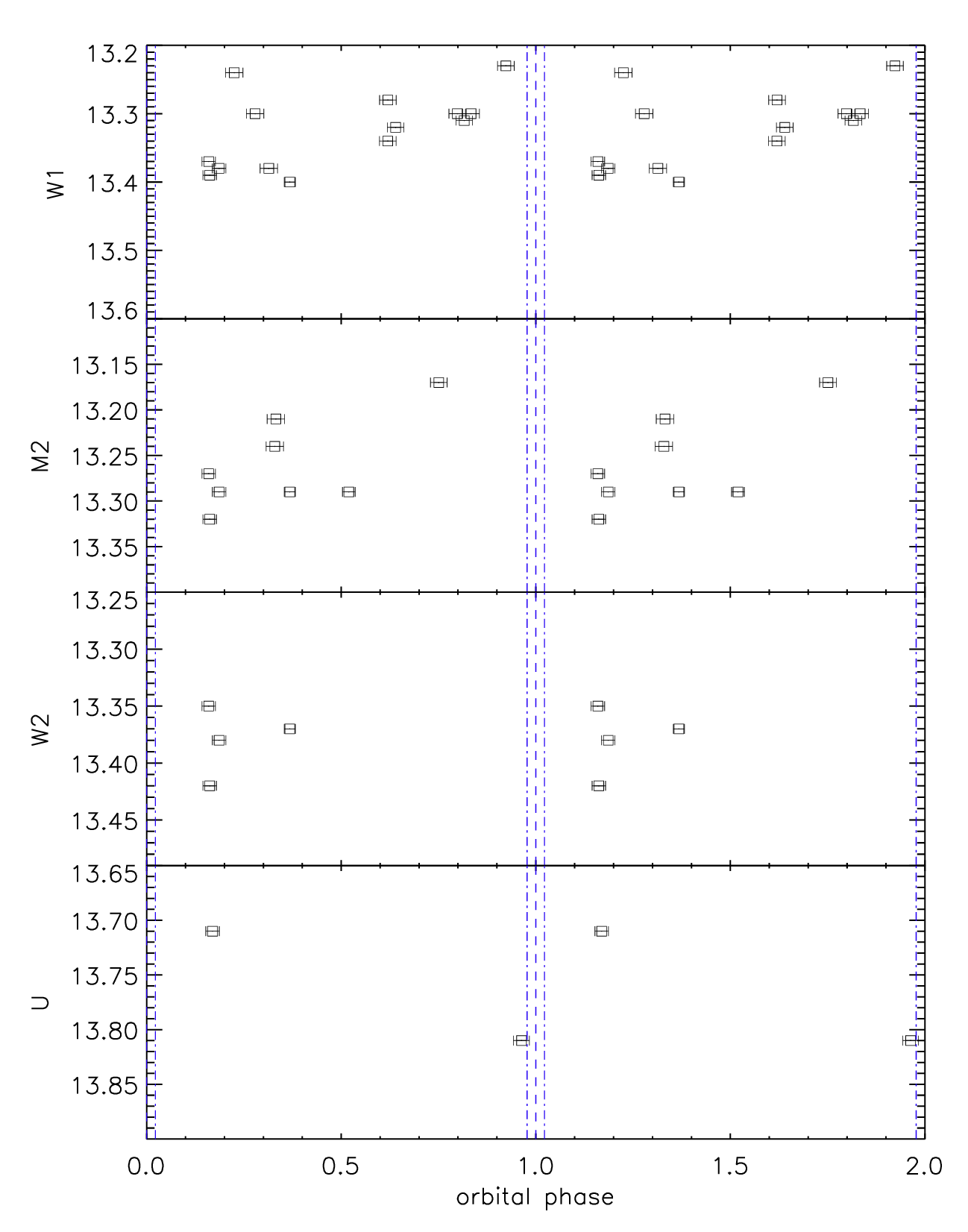}
      \caption{UVOT observations folded at the orbital period of \src. Downward arrows are used to indicate 3$\sigma$ upper limits.
               Vertical dashed lines and dot-dashed lines show the position of the periastron passages
               and their uncertainties, respectively.}
         \label{xray uvot folded}
   \end{figure}

Figure \ref{xray uvot folded} shows the UVOT light curves
in four filters (U: 3465\,\AA; W1: 2600\,\AA; M2: 2246\,\AA; W2: 1928\,\AA)
folded at the orbital period (see Sect. \ref{sect longterm}). Clearly, \src\ is significantly variable
in all bands, with amplitudes $\Delta m_{\rm U}=0.1$, $\Delta m_{\rm W1}=0.16$,
$\Delta m_{\rm M2}=0.15$, and $\Delta m_{\rm W2}=0.07$.
Unfortunately, the data are too sparse to draw any conclusion about the UV variability 
along the orbit of \src.

\begin{table*}
\begin{center}
\caption{Spectral parameters and fluxes of the \xmm, \sw, and \cha\ observations analysed in this work. For observations for which spectral analysis was possible, we reported the unabsorbed luminosity.}
\label{Table spectra}
\resizebox{\columnwidth+\columnwidth}{!}{
\begin{tabular}{lccccccccccc}
\hline
\hline
\multicolumn{12}{c}{\xmm}\\
\hline
\noalign{\smallskip}
    MJD     &    ObsID   &\multicolumn{3}{c}{Exposure [ks]} &         $N_{\rm H}$        &       $\Gamma$           & $\chi^2_\nu$ (d.o.f.) &            absorbed flux           &            unabs. flux           &          $L_{\rm x}$                 & Orb. phase \\
\noalign{\smallskip}
            &            &     pn     &    mos1   &   mos2  &    $10^{22}$\,cm$^{-2}$    &                          &                      &      \multicolumn{2}{c}{(0.3$-$8\,keV) erg\,cm$^{-2}$\,s$^{-1}$}       & (0.3$-$8\,keV) erg\,s$^{-1}$         &            \\
\hline
\noalign{\smallskip}
 51832.63   & 0110000101 &    15.9    &    23.7   &   22.1  &  $0.27{+0.04\atop -0.03}$&      $0.74\pm 0.05$        &  0.892    (115)      &    $9.5\pm0.3\times 10^{-13}$    &$10.4{+0.3\atop-0.3}\times 10^{-13}$ & $4.48{+0.15\atop -0.14}\times 10^{35}$ &   0.082  \\
\noalign{\smallskip} 
 54201.83   & 0404680301 &    13.0    &           &         & $0.44{+0.06 \atop -0.06}$&      $1.03\pm 0.08$        &  1.168  (47)         &    $7.1\pm0.3\times 10^{-13}$    & $8.4{+0.3\atop-0.4}\times 10^{-13}$ & $3.63{+0.14\atop -0.16}\times 10^{35}$ &   0.109 \\
\noalign{\smallskip} 
 55107.21   & 0601211301 &            &           &   31.2  &                          &                            &                      &   $1.7\pm0.5\times 10^{-14}$     &                                    &                                    &   0.413   \\  
\noalign{\smallskip}
\hline 
             &           &            &           &         &                           &                           &                      &                                  &                                    &                                   &            \\
\multicolumn{12}{c}{  }\\
\multicolumn{12}{c}{\emph{Swift}}\\
\hline
\noalign{\smallskip}
    MJD     &    ObsID   &\multicolumn{3}{c}{Exposure [ks]} &         $N_{\rm H}$        &        $\Gamma$          &$C-stat$ (d.o.f.) g-o-f&           absorbed flux           &             unabs. flux           &          $L_{\rm x}$                 & Orb. phase \\
\noalign{\smallskip}
            &            &   \multicolumn{3}{c}{XRT/PC}     &    $10^{22}$\,cm$^{-2}$    &                          &                      &     \multicolumn{2}{c}{(0.3$-$8\,keV) erg\,cm$^{-2}$\,s$^{-1}$}        & (0.3$-$8\,keV) erg\,s$^{-1}$         &         \\
\hline
\noalign{\smallskip}
54696.456  & 00037787001 &    \multicolumn{3}{c}{3.1}      &                           &                           &                      &  $<2.8\times 10^{-13}$  $^*$   &                                    &                                   &   0.368    \\
\noalign{\smallskip}
55003.093  & 00031428001 &    \multicolumn{3}{c}{0.6}      &                           &                           &                      &  $<1.3\times 10^{-12}$  $^*$   &                                    &                                   &   0.148    \\
\noalign{\smallskip}
55542.323  & 00090522001 &    \multicolumn{3}{c}{9.9}      &                           &                           &                      &  $<9.8\times 10^{-14}$  $^*$   &                                    &                                   &   0.520    \\
\noalign{\smallskip}
55607.252  & combined$^5$&    \multicolumn{3}{c}{10.5}     &                           &                           &                      &  $<1.2\times 10^{-13}$  $^*$   &                                    &                                   &   0.685    \\
\noalign{\smallskip}
55681.016  & 00090522005 &    \multicolumn{3}{c}{10.0}     &                           &                           &                      &  $<1.4\times 10^{-13}$  $^*$   &                                    &                                   &   0.872    \\
\noalign{\smallskip}
55742.18   & 00040442001 &    \multicolumn{3}{c}{2.26}      &  $0.33{+0.18 \atop -0.14}$ & $0.68{+0.25\atop -0.24}$ & 93.18 (121); 87.82\% & $8.0{+1.0\atop-0.9}\times10^{-12}$ & $8.8{+1.0\atop-0.9}\times10^{-12}$ & $4.0{+0.5\atop -0.4}\times 10^{36}$ & 0.028  \\
\noalign{\smallskip}
55792.999  & combined$^6$&    \multicolumn{3}{c}{2.5}      &                           &                           &                      &  $<7.7\times 10^{-13}$  $^*$   &                                    &                                   &   0.157    \\
\noalign{\smallskip}

55793.768  & 00040440001 &    \multicolumn{3}{c}{1.6}      &                           &                           &                      &  $9.4 \pm 2.9\times 10^{-13}$ &                                    &                                   &   0.159    \\
\noalign{\smallskip}
55794.905  & 00040440002 &    \multicolumn{3}{c}{0.2}      &                           &                           &                      &  $<2.5\times 10^{-12}$   $^*$  &                                    &                                   &   0.162    \\
\noalign{\smallskip}
55796.30   & combined$^1$&    \multicolumn{3}{c}{21.36}     &  $0.23{+0.12 \atop -0.10}$ & $1.03{+0.23\atop -0.22}$ & 95.89 (113); 97.77\% & $4.1{+0.5\atop-0.5}\times10^{-13}$ & $4.7{+0.5\atop-0.5}\times10^{-13}$ & $2.4{+0.2\atop -0.2}\times 10^{35}$ & 0.166  \\
\noalign{\smallskip}
55797.507  & 00032075002 &    \multicolumn{3}{c}{6.3}      &                           &                           &                      &  $7.5 \pm 1.5\times 10^{-13}$ &                                    &                                   &   0.1687    \\
\noalign{\smallskip}
55797.747  & 00032080001 &    \multicolumn{3}{c}{2.9}      &                           &                           &                      &  $6.5 \pm 1.8\times 10^{-13}$ &                                    &                                   &   0.1693    \\
\noalign{\smallskip}
55798.638  & 00032075003 &    \multicolumn{3}{c}{10.5}     &                           &                           &                      &  $6.4 \pm 1.0\times 10^{-13}$ &                                    &                                   &   0.172    \\
\noalign{\smallskip}
55802.352  & combined$^4$&    \multicolumn{3}{c}{4.1}      &                           &                           &                      &  $7.3 \pm 1.6\times 10^{-13}$ &                                    &                                   &   0.181    \\
\noalign{\smallskip}
55894.422  & 00032194001 &    \multicolumn{3}{c}{0.9}      &                           &                           &                      &  $<7.8\times 10^{-13}$  $^*$   &                                    &                                   &   0.415    \\
\noalign{\smallskip}
57289.163  & 00034071001 &    \multicolumn{3}{c}{0.4}      &                           &                           &                      &  $<1.0\times 10^{-12}$  $^*$   &                                    &                                   &   0.963    \\
\noalign{\smallskip}
57824.949  & 00088083001 &    \multicolumn{3}{c}{7.2}      &                           &                           &                      &  $5.0 \pm 1.1\times 10^{-13}$ &                                    &                                   &   0.326    \\
\noalign{\smallskip}
57825.78   & combined$^2$&    \multicolumn{3}{c}{21.84}     &  $0.29{+1.28 \atop -0.19}$ & $0.77{+1.02\atop -0.31}$ & 58.72 (76); 98.69\%  & $3.5{+0.5\atop-0.5}\times10^{-13}$ & $3.9{+3.0\atop-0.5}\times10^{-13}$ & $1.8{+1.4\atop -0.2}\times 10^{35}$ & 0.328  \\
\noalign{\smallskip}
57826.139  & 00088083002 &    \multicolumn{3}{c}{7.3}      &                           &                           &                      &  $4.1 \pm 1.0\times 10^{-13}$ &                                    &                                   &   0.329    \\
\noalign{\smallskip}
57827.275  & 00088083003 &    \multicolumn{3}{c}{7.3}      &                           &                           &                      &  $4.8 \pm 1.2\times 10^{-13}$ &                                    &                                   &   0.332    \\
\noalign{\smallskip}
\hline
\multicolumn{12}{c}{  }\\
\multicolumn{12}{c}{\cha}\\
\hline
\noalign{\smallskip}
    MJD     &    ObsID   &\multicolumn{3}{c}{Exposure [ks]} &         $N_{\rm H}$        &        $\Gamma$          & $\chi^2_\nu$ (d.o.f.)&           absorbed flux           &             unabs. flux           &          $L_{\rm x}$                 & Orb. phase \\
\noalign{\smallskip}
            &            &\multicolumn{3}{c}{ACIS-S / -I}   &    $10^{22}$\,cm$^{-2}$    &                          &                      &     \multicolumn{2}{c}{(0.3$-$8\,keV) erg\,cm$^{-2}$\,s$^{-1}$}        & (0.3$-$8\,keV) erg\,s$^{-1}$         &         \\
\hline
\noalign{\smallskip}
52549.6     &   2945     &\multicolumn{3}{c}{11.7 (ACIS-S)} &                            &                          &                      &  $9.3{+3.1\atop-2.5}\times10^{-14}$ &                                      &                                     & 0.906  \\
\noalign{\smallskip}
52698.6     &   3907     &\multicolumn{3}{c}{50.8 (ACIS-S)} &                            &                          &                      &  $7.9{+0.6\atop-0.6}\times10^{-14}$ &                                      &                                     & 0.285  \\
\noalign{\smallskip}
54060.5     &   8479     &\multicolumn{3}{c}{42.6 (ACIS-I)} &  $0.41{+0.09 \atop -0.09}$ & $0.85{+0.10\atop -0.10}$ &  1.081 (47)          &$1.16{+0.05\atop-0.05}\times10^{-12}$& $1.37{+0.05\atop-0.05}\times10^{-12}$& $6.3{+0.2\atop -0.2}\times 10^{35}$ & 0.750 \\
\noalign{\smallskip}
54061.8     &   7156     &\multicolumn{3}{c}{39.2 (ACIS-I)} &  $0.48{+0.07 \atop -0.07}$ & $0.87{+0.09\atop -0.08}$ &  1.089 (56)          &$1.23{+0.05\atop-0.05}\times10^{-12}$& $1.43{+0.05\atop-0.05}\times10^{-12}$& $6.6{+0.2\atop -0.2}\times 10^{35}$ & 0.753 \\
\noalign{\smallskip}
54062.6     &   8481     &\multicolumn{3}{c}{16.2 (ACIS-I)} &  $0.47{+0.19 \atop -0.18}$ & $0.91{+0.18\atop -0.17}$ &  0.974 (22)          &$1.44{+0.09\atop-0.09}\times10^{-12}$& $1.68{+0.10\atop-0.09}\times10^{-12}$& $7.7{+8.2\atop -7.3}\times 10^{35}$ & 0.755 \\
\noalign{\smallskip}
55300.9     &   11097    &\multicolumn{3}{c}{29.9 (ACIS-S)} &                            &                          &                      &  $4.4{+0.8\atop-0.7}\times10^{-14}$ &                                      &                                     & 0.905  \\
\noalign{\smallskip}
55307.1     &   11980    &\multicolumn{3}{c}{23.0 (ACIS-S)} &                            &                          &                      &  $5.4{+1.0\atop-0.9}\times10^{-14}$ &                                      &                                     & 0.921  \\
\noalign{\smallskip}
55308.9     &   12200    &\multicolumn{3}{c}{27.1 (ACIS-S)} &                            &                          &                      &  $6.8{+1.0\atop-1.0}\times10^{-14}$ &                                      &                                     & 0.926  \\
\noalign{\smallskip}
55317.7     &   11981    &\multicolumn{3}{c}{34.0 (ACIS-S)} &                            &                          &                      &  $7.0{+1.0\atop-1.0}\times10^{-14}$ &                                      &                                     & 0.948  \\
\noalign{\smallskip}
55318.6     &   12208    &\multicolumn{3}{c}{16.2 (ACIS-S)} &                            &                          &                      &  $1.1{+0.2\atop-0.2}\times10^{-13}$ &                                      &                                     & 0.950  \\
\noalign{\smallskip}
56355.9     &   14674    &\multicolumn{3}{c}{46.5 (ACIS-S)} &                            &                          &                      &       $<2.3\times10^{-13}$ $^*$     &                                      &                                     & 0.589  \\
\noalign{\smallskip}
            &combined$^3$&\multicolumn{3}{c}{239.2 (ACIS-S)}&  $0.73{+0.27 \atop -0.24}$ & $1.63{+0.28\atop -0.26}$ &  1.237 (11)          &  $4.8{+0.4\atop-0.4}\times10^{-14}$ &  $7.6{+1.6\atop-6.0}\times10^{-14}$  &                                     &       \\
\noalign{\smallskip}
\hline
\end{tabular}
}
\end{center}
Notes. $^1$: ObsID of the combined observations: 00040440001, 00032075002, 00032080001, 00032075003.
$^2$: ObsID of the combined observations: 00088083001, 00088083002, 00088083003.
$^3$: ObsID of the combined observations: 3907, 2945, 11097, 11980, 12200, 11981, 12208.
$^4$: ObsID of the combined observations: 00040462002, 00040462003, 00040439002, 00040440003, 00040439003.
$^5$: ObsID of the combined observations: 00090522002, 00090522003, 00090522004.
$^6$: ObsID of the combined observations: 00040462001, 00040439001.
$^*$ 3$\sigma$ upper limit.
\end{table*}

\section{Discussion}
\label{sect. discussion}

\subsection{Secular spin-up}
\label{sect. secular spin-up}

Our data analysis of \rxte\ data led to a lower number of detections
compared to the works of \citet{Klus14} and \citet{Yang17}.
This is due to the highest threshold of detection adopted in our work (99.9\% versus 99\%)
and to the different data analysis reduction (for the sophisticated data analysis procedure adopted by
\citet{Klus14} and \citet{Yang17}, see \citealt{Galache08}).
Since we set a higher detection threshold in the Lomb--Scargle analysis,
we  consider only the strongest signals. Our measurements are thus
less contaminated by other signals (e.g. from other pulsars of the SMC in the field of view of \src) which could lead to lower 
precision and accuracy of the measurements.
The improvement obtained by the increase in the detection threshold
can be seen in Fig \ref{spin evol} where the values of the spin periods
are less scattered than they are in  the plots of \citet{Klus14} (figure B39)
and \citet{Yang17}\footnote{See \url{https://authortools.aas.org/AAS03548/FS9/figset.html}}.

\src\ shows a long-term spin-up rate of $\dot{P}=(-3.00\pm0.12)\times 10^{-3}$\,s\,day$^{-1}$,
which is more precise than the spin-up rate reported in \citet{Yang17}
[$\dot{P}=(-4.1\pm1.6)\times 10^{-3}$\,s\,day$^{-1}$].
Such high secular spin-up, when compared with those of other accreting pulsars, 
indicates that the pulsar of \src\ is likely far from its equilibrium period.
For comparison, the Be/XRB SAX\,J2103.5+4545 has shown, since its discovery in 1997, 
a secular spin-up of $\dot{P}\approx -1.2\times 10^{-3}$\,s\,day$^{-1}$ \citep{Camero14}
and it is believed that it has not yet reached its equilibrium period \citep{Baykal02}.
If the pulsar of \src\ were in equilibrium,
  it would be possible to obtain an estimate of its magnetic field strength (see e.g. \citealt{Klus14}).
  Since \src\ is not in equilibrium, we can just calculate an upper limit
  for its magnetic field.
  For this purpose, we used the calculations presented in \citet{Ikhsanov07}
  to show that slow pulsars ($P_{\rm spin} \sim 10^2-10^4$\,s) in equilibrium
  do not need supercritical initial magnetic fields of the NS ($B\geq B_{\rm cr} \approx 4.4 \times 10^{13}$\,G).
  Adapting the calculations presented for the spherical accretion scenario 
(see equation 8 in \citealt{Ikhsanov07}),
  the maximum magnetic field of \src\ would be
\begin{eqnarray} 
  B_{\rm sph} &\lesssim&  2.7\times 10^{10} P_{\rm eq} k_{\rm t}^{-1/2} \xi^{1/2}_{0.2} M_{1.4} L_{37}^{1/2} \times \nonumber \\
            & \times & \left ( \frac{v_{\rm rel}}{400\mbox{\,km\,s}^{-1}} \right )^{-2} \left ( \frac{P_{\rm orb}}{250\mbox{\,d}} \right )^{-1/2} \approx 3\times 10^{12} \mbox{ G,} \label{Bsph eq}
\end{eqnarray}
where $k_{\rm t}$ is a parameter of the order of unity (\citealt{Ikhsanov07}; \citealt{Ikhsanov02}),
$M_{1.4}=M_{\rm NS}/(1.4M_\odot)$ is the mass of the NS, $L_{37}=$ is the X-ray luminosity in units of $10^{37}$\,erg\,s$^{-1}$,
$v_{\rm rel}$ is the relative velocity between the NS and the wind from the companion star, and
$\xi_{0.2}=\xi/0.2$ is a factor that takes into account the reduction of the angular momentum accretion rate
caused by velocity and density inhomogeneities in the accretion flow (see \citealt{Ikhsanov07} and references therein).
Since the pulsar of \src\ is not in equilibrium,
we calculated the upper limit $B_{\rm sph}$ for the conservative case of $P_{\rm eq}=750$\,s,
and we assumed an average luminosity of $\sim 4 \times 10^{35}$\,erg\,s$^{-1}$.
We obtained it from the \sw, \cha, \xmm, and \asca\
observations. We did not consider
the \rxte\ observations because they would introduce a bias due to the relatively low sensitivity
of PCA compared to the other three instruments.

\citet{Ikhsanov07} pointed out that the probability of observing in X-ray
  a long-period accreting pulsar fed by an accretion disc at a rate of $\sim 10^{15}$\,g\,s$^{-1}$
  is very low because the accretion disc would spin up  the pulsar at a high rate,
  implying a lifetime of the pulsar during the accretion phase
  of $\ll 1000$\,yr, i.e. several orders of magnitude smaller than the typical
  lifetime of accreting NSs in high-mass X-ray binaries.
  Therefore, according to \citet{Ikhsanov07}, the accretion disc scenario
  is unlikely for binary systems similar to \src.
  Nonetheless, we considered here, for completeness,
  also this case.
  From equation 7 in \citet{Ikhsanov07}, the upper limit of the
  magnetic field  in the accretion disc case would be
\begin{equation} \label{B eq}
B_{\rm disc} \lesssim 4.2 \times 10^{11} P_{\rm eq}^{7/6} \kappa_{0.5}^{7/24} k_{\rm t}^{-7/12} M_{1.4}^{1/3} L_{37}^{1/2} \approx 2\times 10^{14} \mbox{ G,}
\end{equation}
where $\kappa_{0.5}=\kappa/0.5$ is a parameter that takes into account the geometry
of the accretion flow ($\kappa=0.5$ corresponds to disc geometry, while $\kappa=1$ to the spherical geometry).
We assumed the same $P_{\rm eq}$ and $L_{37}$ of the spherical accretion case.

\subsection{Long-term variability}
\label{subsect disc long-term}

\src\ (Figs. \ref{xray ogle} and \ref{xray folded}) 
shows an enhanced X-ray luminosity
($\approx 10^{36}-10^{37}$\,erg\,s$^{-1}$) at periastron.
In a few cases, the source has been detected at high X-ray
luminosities (similar to those observed at periastron) across the entire orbit,
including orbital phases near to the apastron.
Moreover, \src\ shows a high variability far from periastron,
with a maximum variability factor of $\sim 270$.

\begin{figure}
   \centering
   \includegraphics[width=\columnwidth]{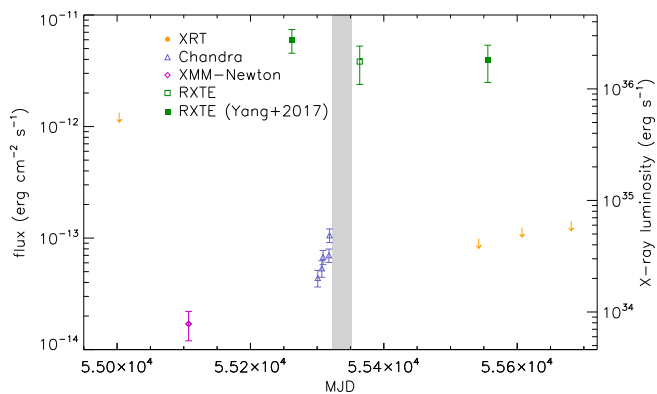}
      \caption{X-ray light curve of \src\ around the periastron passage at $\approx 55340$\,MJD,
        showing the \rxte\ detections reported in \citet{Yang17} and \citet{Klus14} (filled green boxes).
        The grey vertical stripe shows the time of periastron passage
               according to the ephemeris calculated by \citet{Schmidtke13}.}
         \label{xray zoom}
   \end{figure}

The periodic outbursts at periastron displayed by \src\ are consistent with
the definitions of type I outburst given in Sect. \ref{sect intro}.
In the framework of the truncation disc model, the system
is expected to have a high eccentricity to show regular type I outbursts.
On the other hand, the strong X-ray variability and high luminosity states
observed out of periastron cannot be reconciled with the typical variability
of Be/XRBs, traditionally described in terms of type I and type II outbursts.
The same conclusion was reached by \citet{Coe04} on the basis of three \asca\
outbursts ($L_{\rm x} \approx 10^{35}$\,erg\,s$^{-1}$) observed far from periastron. 
The anomalous X-ray variability of \src\ noted by \citet{Coe04} is 
confirmed by the results presented in Sect. \ref{sect. results};
 it is further complicated when the \rxte\ detections reported in \citet{Yang17}
and \citet{Klus14} are taken into account.
Figure \ref{xray zoom} shows two interesting \rxte\ detections
at $t_{\rm a}=55262.10$\,MJD and $t_{\rm b}=55556.02$\,MJD (\citealt{Yang17}; \citealt{Klus14}).
Together with the \cha\ and \sw\ detections and upper limits 
reported in Table \ref{Table spectra} and shown 
in Fig. \ref{xray zoom}, they constitute two peculiar events
characterized by high variability 
and with timescales $t \ll P_{\rm orb}$, located far from periastron.
In particular, from $\sim$55262\,MJD (\rxte\ observation) to 
$\sim$55300\,MJD (\cha\ observations), about 90 days before the periastron passage ($\phi_{\rm orb} \approx 0.8-0.9$),
the X-ray flux decreased by a factor of $\sim 150$ in less than about 38 days.
\cha\ observed the source five times in the subsequent 18 days. 
During this period, \src\ showed a slow increase in flux, 
from $\approx 4\times 10^{-14}$\,erg\,cm$^{-2}$\,s$^{-1}$ to $\approx 10^{-13}$\,erg\,cm$^{-2}$\,s$^{-1}$.
After this X-ray dip caught by \cha, \src\ was observed again by \rxte,
close to periastron ($\sim$55364\,MJD), with a flux of $\approx 4.3 \times 10^{-12}$\,erg\,cm$^{-2}$\,s$^{-1}$.
Another jump in luminosity was observed in the subsequent orbital cycle,
when \src\ showed an increase in brightness 
from  $F_{\rm x} \leq 10^{-13}$\,erg\,cm$^{-2}$\,s$^{-1}$ at $\sim$55542\,MJD (\sw/XRT)
to $F_{\rm x} \approx 4 \times 10^{-12}$\,erg\,cm$^{-2}$\,s$^{-1}$ at $\sim$55556\,MJD (\emph{RXTE}, \citealt{Yang17}),
and then a decrease to $F_{\rm x} \leq 10^{-13}$\,erg\,cm$^{-2}$\,s$^{-1}$ at $\sim$55607\,MJD (\sw/XRT).
The peak luminosity was observed close to apastron (at the orbital phase $\sim 0.56$).
The timescale of the variability and the flux levels are similar
to those of the previous event.
This variability could be ascribed to multiple short-term
outbursts peaking randomly in the orbital phase.
Hereafter, we consider two possible mechanisms to explain the observed variability.

   \subsubsection{Gating mechanism}
   \label{sect. gating mechanism}

The decrease in luminosity observed  during the events reported in Sect. \ref{sect. results}
might be caused by the onset of the centrifugal barrier when the mass
inflow rate decreases to a certain limiting value. Transitions from direct
accretion to centrifugal inhibition of accretion 
(or propeller, \citealt{Illarionov75}) were proposed
by \citet{Stella86} to explain the high dynamical range, spanning up to six
orders of magnitudes, of the X-ray luminosity of some transient Be/XRBs.
These transitions depend on the amount of inflowing matter, on the magnetic field strength
of the NS and on its spin period. 
They can be easily understood by introducing
the definitions of corotation and magnetospheric radii.
The corotation radius ($r_{\rm c}$) is the distance from the NS at which there is a balance
between the NS angular velocity and the Keplerian angular velocity.
For \src\ we obtain 
$r_{\rm c} = (GM_{\rm NS})^{1/3} (P_{\rm s}/2\pi)^{2/3} \approx 1.4 \times 10^{10}$\,cm,
where we assumed $M_{\rm NS}=1.4$\,M$_\odot$ and $P_{\rm s}=750$\,s.
The magnetospheric radius ($r_{\rm m}$) is the distance from the NS 
where the magnetic field pressure equals the ram pressure 
of the accreted ionized plasma dragged along the field lines.
If the magnetospheric radius is pushed inside the corotation radius by 
a sufficiently high amount of inflowing
matter, the matter is expected to be effectively accreted by the NS.
If the inflowing matter decreases, for example because of small variations in the wind
properties around the NS (density and wind velocity), the magnetospheric radius can 
expand beyond the corotation radius. 
In this case, the plasma dragged by the magnetic field lines
at the magnetosphere would rotate at a super-Keplerian rate,
the magnetosphere would behave as a closed barrier, and direct accretion would be
inhibited. According to \citet{Stella86}, the minimum X-ray luminosity caused by
direct accretion is obtained assuming $r_{\rm m} = r_{\rm c}$, and it is given by
\begin{equation} \label{eq stella}
L_{\rm min} = \frac{GM_{\rm NS}\dot{M}_{\rm min}}{R} \approx 4\times 10^{30} R_6^{-1} M_{1.4}^{-2/3} \mu^2_{30} P_{750}^{-7/3} \mbox{ erg\,s}^{-1}
,\end{equation}
where $\dot{M}_{\rm min}$ is the minimum mass accretion rate for which
the NS accretes directly, 
$R_6=10^6$\,cm is the radius of the NS, $M_{1.4}=1.4$\,M$_\odot$ is the mass of the neutron star, 
$\mu_{30}=10^{30}$\,G\,cm$^3$ is the NS magnetic moment, and $P_{750}=750$\,s
is the spin period of \src.
In Eq. (\ref{eq stella}) we assumed a 100 \%\ efficient conversion of the 
gravitational potential energy of the accreted matter into X-ray radiation.
If we assume that the minimum X-ray luminosity caused by direct accretion is the
lowest luminosity observed by \xmm\ and \cha\ when the pulsation was detected \citep{Hong17, Haberl08},
i.e. $L_{\rm min} \approx 3\times 10^{35}$\,erg\,s$^{-1}$, from Eq. (\ref{eq stella})
we find that the pulsar of \src\ should have 
a magnetic field strength of $\approx 2.7 \times 10^{14}$\,G.
When the centrifugal barrier becomes active ($r_{\rm m} = r_{\rm c}$),
the maximum allowed luminosity in the propeller regime is given by (see \citealt{Campana97})
\begin{equation} \label{eq campa}
L_{\rm max} = \frac{GM_{\rm NS}\dot{M}_{\rm min}}{r_{\rm c}} \approx 2.3 \times 10^{27} \mu_{30}^2 M_{1.4}^{-1} P_{750}^{-3} \mbox{ erg\,s}^{-1}
,\end{equation}
which is of the order of $10^{30}$\,erg\,s$^{-1}$ when we assume
$B \approx 2.7 \times 10^{14}$\,G\footnote{The approximated equation 2
for the calculation of the magnetospheric radius in \citet{Stella86} might lead
to overestimating it, as discussed in \citet{Bozzo09}.
Therefore, the values given in Eqs. (\ref{eq stella}) and (\ref{eq campa})
have to be taken with caution, and could be affected by an uncertainty of a factor of a few.}. 
The value of $L_{\rm max}$ found with Eq. \ref{eq campa}
is more than three orders of magnitude
lower than the lowest luminosity observed by \xmm\ in October 2009,
$L_{\rm lowest}=8\times 10^{33}$\,erg\,s$^{-1}$. This value of 
$L_{\rm lowest}$ is about 10--100 times higher than the X-ray luminosity
of the brightest known Be stars \citep{Motch07}, and about $10^3$ times higher
than the X-ray luminosity of typical Be stars. 
Therefore, the source of this emission is most likely the NS rather than the Be star.
The lack of detection of pulsations in the \cha\ observations
at low luminosity (Sect. \ref{sect. results}) might be due to insufficient statistics.
To verify this possibility, we performed a timing analysis 
of the \cha\ data at the lowest flux (55300.9$\leq t\leq$55318.6\,MJD)
which was not  reported in previous works.
We used barycentred background subtracted light curves, assuming two different bin sizes: 50\,s and 100\,s.
We searched for periodicities using the Lomb--Scargle periodogram technique 
as described in Sect. \ref{subs timing}.
We did not detect any significant ($\geq 3\sigma$) pulsating signal in the data.
We performed simulations on the two light curves built with different bin sizes (50\,s and 100\,s)
to set a 3$\sigma$ upper limit on the pulsed fraction of a sinusoidal signal of about $45$\,\%.
This upper limit is comparable with the typical pulsed fraction of the source 
detected at higher luminosities (see e.g. \citealt{Hong17}).
Therefore, we conclude that the transitions from centrifugal inhibition of accretion to direct accretion
is unlikely to be responsible for the observed
jumps in X-ray luminosity.

\subsubsection{Accretion from a cold disc}

Another possibility is given by the scenario proposed by \citet{Tsygankov17}.
They showed that, after an outburst, a sufficiently slow pulsar in a Be/XRB
could switch to an accretion state in which the pulsar is fed by a cold accretion disc.
This accretion state is possible if the mass accretion rate is sufficiently high
to open the centrifugal barrier ($r_{\rm m}<r_{\rm c}$)
and sufficiently low to have an accretion disc colder than $\approx 6500$\,K.
The latter condition is verified if\footnote{\citet{Tsygankov17} presented accurate conditions 
to have accretion from a cold disc (see their equations 12 and 13).
However, due to the lack of enough information for \src, we use here the simplified
conditions, based on the assumption that the temperature of the accretion disc reaches its 
maximum at the inner radius (see equations 6 and 7 in \citealt{Tsygankov17}).}
\begin{equation} \nonumber
L < L^{\rm cold} = 9\times 10^{33} \kappa^{1.5} M_{1.4}^{0.28}R_6^{1.57}B_{12}^{0.86} \mbox{ erg\,s}^{-1} \mbox{ ,}
\end{equation}
where $B_{12}=B/(10^{12}$\,G).
The magnetic field strength of the pulsar in \src\ is not known.
Assuming four test values, $B_{12}=10^{12}$, $B_{13}=10^{13}$, $B_{14}=10^{14}$\,G, and $B_{15}=10^{15}$\,G,
we find $L^{\rm cold}_{12}=3\times 10^{33}$\,erg\,s$^{-1}$, 
$L^{\rm cold}_{13}=2.3\times 10^{34}$\,erg\,s$^{-1}$,
$L^{\rm cold}_{14}=1.7\times 10^{35}$\,erg\,s$^{-1}$,
and $L^{\rm cold}_{15}=1.2\times 10^{36}$\,erg\,s$^{-1}$.
\citet{Tsygankov17} also noted that the accretion state (cold disc versus propeller)
after an outburst is determined by the spin period and magnetic field strength.
Stable accretion from a cold disc, instead of the onset of the propeller regime, is possible if
\begin{equation} \nonumber
P_{\rm spin} > P^* = 36.6 \kappa^{6/7}B_{12}^{0.49}M_{1.4}^{-0.17}R_6^{1.22} \mbox{ s.}
\end{equation}
Assuming again four possible values for the magnetic field strength of the pulsar,
we find $P^*_{12}=20$\,s, $P^*_{13}=62$\,s, $P^*_{14}=193$\,s, and $P^*_{15}=596$\,s.
Therefore, the relatively high luminosities of \src\ observed far from periastron 
($L_{\rm x}\approx 2\times 10^{35}-10^{36}$\,erg\,s$^{-1}$) might 
be caused by stable accretion from a cold accretion disc only if 
the magnetic field of the pulsar is high, i.e. $B\gtrsim 10^{14}$\,G.
Although magnetic fields of this magnitude are possible in accreting NSs
in high-mass X-ray binaries,
they should represent rare cases compared to the typical values measured so far
($B\approx 10^{11-13}$\,G, \citealt{Revnivtsev15}).
Therefore, accretion from a cold disc is unlikely
to be the reason for the short-term variability at random orbital phases in \src.

\subsubsection{Perturbed circumstellar disc}

If we exclude that the two previous mechanisms  
are  responsible for the high X-ray variability
of \src, one possibility is that
outside  the time periods with enhanced X-ray activity ($L_{\rm x}\geq 10^{35}$\,erg\,s$^{-1}$),
\src\ is a persistent emitter, with luminosity of the order of  $L_{\rm lowest}$.
There is a small subclass of Be/XRBs that display persistent X-ray emission at low
luminosity levels ($L_{\rm x}\geq 10^{34}-10^{35}$\,erg\,s$^{-1}$)
and low X-ray variability (compared to Be/XRB transients; see \citealt{Reig99} and references therein).
\src\ has some properties in common with the class of persistent Be/XRBs, which also contains
slowly rotating NSs ($P_{\rm spin} > 200$\,s) in wide orbits ($P_{\rm orb} > 30$\,d; \citealt{Reig07,Reig11}).
On the other hand, persistent Be/XRBs have low eccentricities ($e<0.2$),
at odds with the high eccentricity inferred for \src\ 
to explain the outbursts at periastron (see beginning of this section).
Nonetheless, we note that another persistent Be/XRB, RX\,J0440.9+4431 
($P_{\rm orb}\approx 150$\,d, $P_{\rm spin}\approx 205$\,s), rarely showed flaring activity
at periastron (peak luminosity $L_{\rm x}\approx 7.5\times 10^{36}$\,erg\,s$^{-1}$,
2$-$30\,keV; \citealt{Ferrigno13}). 
In this system, the persistent emission ($L_{\rm x}\approx 5\times 10^{34}$\,erg\,s$^{-1}$) 
might be a consequence of the accretion
of the rarefied wind produced by the companion star outside  the circumstellar disc,
while the flares are related to the accretion at periastron 
of the dense wind of the circumstellar disc \citep{Ferrigno13}.

The observed jumps in luminosity might suggest that the material accreted by the NS
across the orbit is strongly inhomogeneous, 
and it is characterized by strong variations in density and stellar wind velocity
uncorrelated with the orbital phase.
In Sect. \ref{sect. results} we showed that the variability timescales of these jumps
is $\lesssim 37$\,d. Assuming a circular orbit, an orbital period of $393$\,d,
and masses of the two stars of $M_{\rm *}=18$\,M$_\odot$ \citep{Klus14}
and $M_{\rm NS}=1.4$\,M$_\odot$,
we find that the sizes of the structures responsible for the variability are 
of the order of $2.5\times10^{13}$\,cm.
Such large structures encountered by the NS along the orbit
might be the result of the tidal interaction of the NS with the circumstellar
disc during previous orbital passages.
As suggested by \citet{Okazaki01}, in binary systems with very long orbital periods
($\gtrsim 200$\,d), the Be disc might be able to spread out 
significantly beyond the truncation radius while the NS is far from periastron. 
Moreover, three-dimensional hydrodynamic simulations have shown  that in a systems with a 
misaligned orbital plane and a circumstellar disc, a warped and eccentric circumstellar disc
can develop \citep{Martin14}.
We propose that  the disc expansion in \src\ might occur not uniformly on the equatorial plane of the Be star,
and when the NS crosses the circumstellar disc again, it might accrete large
isolated  structures characterized by high density and low wind velocity.
%
It is worth noting that GRO\,J1008$-$57, another Be/XRB with a long
orbital period ($\sim 250$\,d) showed an anomalous variability in 2014/2015,
namely three outbursts in a single orbit, with the peak of the third  reached
at apastron \citep{Kuehnel17}.
\citet{Kuehnel17} discussed the peculiar variability of GRO\,J1008$-$57
in the framework of misaligned orbital plane and circumstellar disc, 
with outbursts occurring at the intersection between these two planes.
We point out that variability displayed by GRO\,J1008$-$57 in 2014/2015,
shows some similarities with that observed in \src\ and presented in this work.
Nonetheless, we note that it is difficult to explain the X-ray variability of \src\
with the scenario proposed for GRO\,J1008$-$57. 
In fact, the high luminosity states of \src\ observed 
far from periastron by \asca, \cha, and \rxte, always occur 
at different orbital phases, while the other intersection (at periastron)
is constant in phase with time.
Such strong variability of the phase at which one of the two intersections occurs
would require a warped/tilted or high precessing circumstellar disc.

\section{Conclusions and future work}

We presented an analysis of archival \sw,
\cha, \xmm, \rxte, and \inte\ data of the Be/XRB \src.
The spectral analysis shows an anti-correlation between the power law slope
describing the X-ray continuum and the X-ray flux.
This behaviour has been observed in other accreting pulsars
in early-type systems (e.g. \citealt{Reig13,Romano14,Malacaria15}).

\src\ shows a secular spin-up of $\dot{P}=(-3.00\pm0.12)\times 10^{-3}$\,s\,day$^{-1}$,
which suggests that the pulsar has not yet achieved the equilibrium period.

To gain more information about the X-ray properties of 
\src, we studied its long-term X-ray variability,
making use of \asca, \sw,
\cha, \xmm, \rxte, and \inte\ data analysed here and in other works.
We found that \src\ shows a high X-ray variability, with
high luminosity states ($L_{\rm x} > 5\times 10^{35}$\,erg\,s$^{-1}$)
caught by \cha, \sw, \rxte, and \asca\ far from periastron,
which suggests that the NS experienced prolonged periods of relatively
high accretion rate in different orbital cycles,
likely due to the presence of a stable ($t\gtrsim 1000$\,d) extended circumstellar disc.
Two \rxte\ detections reported by \citet{Yang17} and \citet{Klus14},
together with \cha\ and \sw\ data analysed in this work,
would indicate two cases of anomalous fast variability far from periastron.
If the \rxte\ detections reported by \citet{Yang17} and \citet{Klus14} are not due to spurious effects introduced by statistical fluctuations,
we showed that the observed anomalous fast variability discussed in Sect. \ref{sect longterm} is
likely due to complicated tidal interactions
of the NS with an extended circumstellar disc.
It would be important to observe  these flaring events again with future
observations. The hypotheses proposed here and in other papers \citep{Coe04}
to explain the long-term variability of \src\
could be verified through a more continuous X-ray monitoring 
of the source, coupled with
simultaneous spectroscopic observations in optical/UV, especially focused on the
study of the long-term variability of emission lines such as H$\alpha$.
This would provide important information about possible changes of the properties
of the circumstellar disc, such as its size and orientation.

\begin{acknowledgements}
We thank the anonymous referee for the useful comments that improved the manuscript.
L.D. thanks Jun Yang for answering some questions about the library of X-ray pulsars in SMC.  
This paper is based on
data from observations with, XMM-Newton, Swift, and Chandra X-ray observatory.
XMM-Newton is an ESA science mission with instruments 
and contributions directly funded by ESA Member States and NASA.
Chandra data were obtained from the Chandra Data Archive.
This paper is based on data from observations with INTEGRAL, 
an ESA project with instruments and science data centre funded by ESA
member states (especially the PI countries: Denmark, France, Germany,
Italy, Spain, and Switzerland), Czech Republic, and Poland,
and with the participation of Russia and the USA.
This work is supported by the Bundesministerium f\"ur
Wirtschaft und Technologie through the Deutsches Zentrum f\"ur Luft
und Raumfahrt (grant FKZ 50 OG 1602).
P.R. acknowledges contract ASI-INAF I/004/11/0.
\end{acknowledgements}

\bibliographystyle{aa} 
\bibliography{ld0049}

\end{document}